\documentclass[a4paper]{jpconf}
\usepackage{graphicx}
\usepackage{epsfig}
\begin{document}
\title{Mass and rapidity dependent top quark forward-backward asymmetry in the effective Lagrangian approach}

\author{Dong-Won Jung,$^1$ P. Ko,$^2$ and
Jae Sik Lee$^3$}

\address{$^1$$^{,2}$ School of Physics, KIAS, Seoul 130-722, Republic of Korea \\
$^3$ Department of Physics, Chonnam National University,
300 Yongbong-dong, Buk-gu, Gwangju, 500-757, Republic of Korea}

\ead{$^1$dwjung@kias.re.kr, $^2$pko@kias.re.kr, and $^3$jslee@jnu.ac.kr}

\begin{abstract}
We study the invariant mass and rapidity dependent top quark forward-backward 
asymmetry from the effective Lagrangian viewpoint. The Wilson coefficients 
are constrained by the experimental observations and 
the concrete models that reproduce the 
low energy effective Lagaragians are considered. Some of them are disfavored 
and others relatively favored. For each cases, we estimate the appropriacy of 
the effective Lagrangian approach. 
\end{abstract}

\section{Introduction}
A few years ago, the Tevatron collabroration reported the top forward-backward (FB) 
asymmetry ($A_{\rm FB}$) and since then it has been drawing much interest from high 
energy physics community. In addtion to the integrated asymmetry of 
$A_{\rm FB}({\rm CDF})  = 0.158 \pm 0.074$ \cite{ljet} 
\footnote{The most recent values are 
$
A_{\rm FB} ({\rm CDF}) = 0.162 \pm 0.042 ~~\cite{Aaltonen:2012it},~~~ 
A_{\rm FB} ({\rm D0}) = 0.196 \pm 0.065 ~~\cite{Abazov:2011rq}.
$ There are still around 2-$\sigma$ deviation from the standard 
model calculation.}
 , the mass and rapidity 
dependent asymmetries are also reported and the most recent results are 
given by the CDF collaboration \cite{cdfljetsnew}. 

In the previous publications, we performed the detailed analyses 
on the integrated asymmetry with the effective Lagarangian approach 
\cite{Jung:2009pi, Jung:2010yn} and also on the rapidity and invariant 
mass dependent asymmetries \cite{Jung:2011ym}. In this paper, we extend 
our analysis in Ref. \cite{Jung:2011ym} with updated data, also including the 
consideration of more general cases and their validities. 

\section{Effective Lagarangian}
If the new physis scale is higher than the energy scale considered, we
 can adopt the effective Lagrangian to describe the low energy phenomena. 
Since no resonance has been observed at the Tevatron energy scale, we adopted 
the effective Lagrangian approach to draw the information on the new physics 
which might be responsible for top quark $A_{\rm FB}$. Most generally we can 
write down the dimension-6 operators for 
$q\bar{q} \rightarrow t \bar{t}$~\cite{Jung:2009pi,Jung:2010yn}:
\begin{eqnarray}
\mathcal{L}_6 &=& \frac{g_s^2}{\Lambda^2}\sum_{A,B}
\left[
C^{AB}_{8q}(\bar{q}_A T^a\gamma_\mu q_A)(\bar{t}_B T^a\gamma^\mu  t_B)\right]\,.
\end{eqnarray}
With shorthand notations of 
$C_1 \equiv C_{8q}^{LL}+C_{8q}^{RR}$ and  $C_2 \equiv C_{8q}^{LR}+C_{8q}^{RL}$,
$A_{\rm FB}$ must be expanded with $\sigma_{NLO}^{int}, \sigma_{NP}^{int}.$ In that case, 
the FB asymmetry must be written 
\begin{eqnarray}
A_{FB} \simeq A_{FB}^{SM}+ \Delta \sigma_{NP}^{int}/\sigma^{SM}_0.
\end{eqnarray} Then the deviation from the $SM$ are expressed as 
\begin{eqnarray}
\label{eq:bsm}
\Delta\sigma_{t\bar{t}}&\equiv&
\sigma_{t\bar{t}} - \sigma_{t\bar{t}}^{\rm SM} \propto ( C_1 + C_2 ),  \\
\Delta A_{\rm FB}&\equiv&
A_{\rm FB}-A_{\rm FB}^{\rm SM} \propto (C_1 - C_2). 
\end{eqnarray}
We will use $A_{\rm FB}({\rm CDF})  = 0.158 \pm 0.074$ in order to fix the effective couplings and predict the $M_{t\bar{t}}$ and $\Delta y$ dependent $A_{\rm FB}$ 
for those $C_i$'s within $1$-$\sigma$ range. 
We found that $C_1\,(1\,{\rm TeV}/\Lambda)^2$ and $-C_2\,(1\,{\rm TeV}/\Lambda)^2$ 
take values between $\sim -0.5$ and $\sim 2.5$, 
see Fig.~1 in Ref.~\cite{Jung:2010yn} for updated results.

\section{FB asymmetries as functions of invariant mass $M_{t\bar{t}}$ and 
rapidity difference $\Delta y$}
It will be instructive to start this section with old wisdom of electroweak interaction. 
The FB asymmetry of the muon pair was measured at {\rm PETRA} \cite{Wu:1984ik}, the 
center of mass energy was far below the $Z^0$ pole mass. One can observe clear FB 
asymmetry by the parity violating weak interaction, which is absent in QED. More 
explicitly, $A_{\rm FB}$ of muon pair far below the $Z^0$ mass can be approximated as
\begin{equation}
A_{FB} (s)  \simeq
 -\frac{3 G_F}{\sqrt{2}} ~\frac{s}{4\pi \alpha}~ (g_L - g_R)^2 
\equiv k G_F s ,  
\end{equation}
which is negative definite, a generic feature of the new vector boson with universal
couplings to the initial and the final fermions and antifermions.
The PETRA measurement of $A_{FB} (s)$ in the region far below the $Z^0$ pole
is that the $A_{\rm FB} (1200 {\rm GeV}^2) \simeq - 0.1$, which can be 
translated into 
\[
k = - 7.18  ,
\] 
compared with the SM prediction: $k = -5.78$. Note that we can get the rough 
size of $k$ (or $(g_L - g_R)^2 / M_Z^2$) 
only from the interference term between the  
QED photon and the $Z^0$ boson exchanges in the limit $s\rightarrow 0$
(near threshold), if $s \ll M_Z^2$. 

Similar things can be done for $q\bar{q} \rightarrow t\bar{t}$ process. In this case with 
the dimension-6 operator the asymmetry is
\begin{equation}
\widehat{A}_{\rm FB} ( M_{t\bar{t}} ) = 
\frac{\hat{\beta}_t \frac{\hat{s}}{\Lambda^2} 
(C_1 - C_2) }{
 \frac{8}{3} 
 \left[1 + \frac{\hat{s}}{2 \Lambda^2}  (C_1 + C_2) \right] 
+\frac{16 \hat{s}}{3 m_t^2} 
 \left[ 1+ \frac{\hat{s}}{2 \Lambda^2} (C_1 + C_2) \right] 
}
\simeq \frac{3 \hat{\beta}_t \frac{\hat{s}}{\Lambda^2} 
(C_1 - C_2) }{8 + 16 \frac{\hat{s}}{m_t^2}} .
\end{equation}
In any case, the whole point is that the FB asymmetry near the threshold 
is approximately linear in $\hat{s}$ modulated by 
$\hat{\beta}_t = \sqrt{1 - 4 m_t^2/ \hat{s}}$ 
with a small slope parameter that could have either sign depending on 
$(C_1 - C_2)$,  namely the underlying new physics affecting 
$q\bar{q} \rightarrow t \bar{t}$. 

\begin{figure}[!t]
\begin{center}
\vspace{-0.0cm}
\begin{tabular}{cc}
{\epsfig{figure=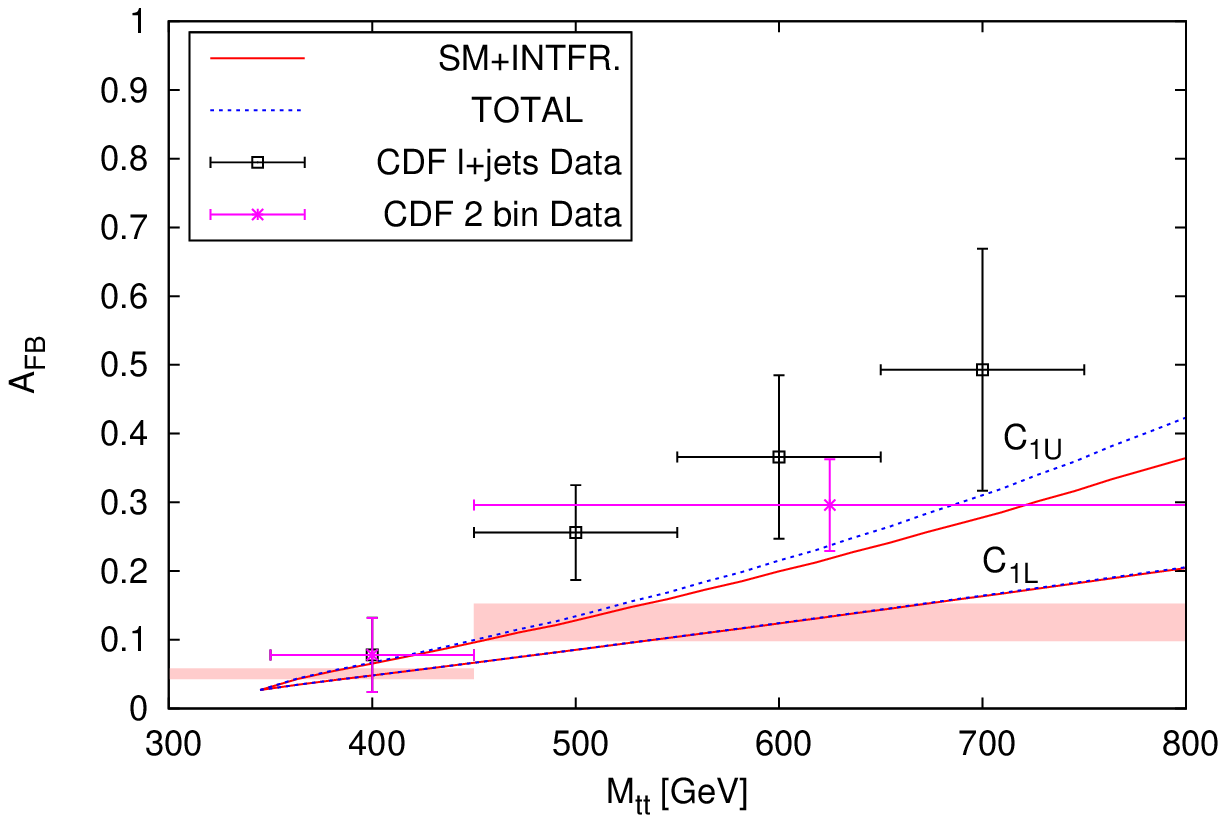,height=4.0cm,width=7.0cm}} &
{\epsfig{figure=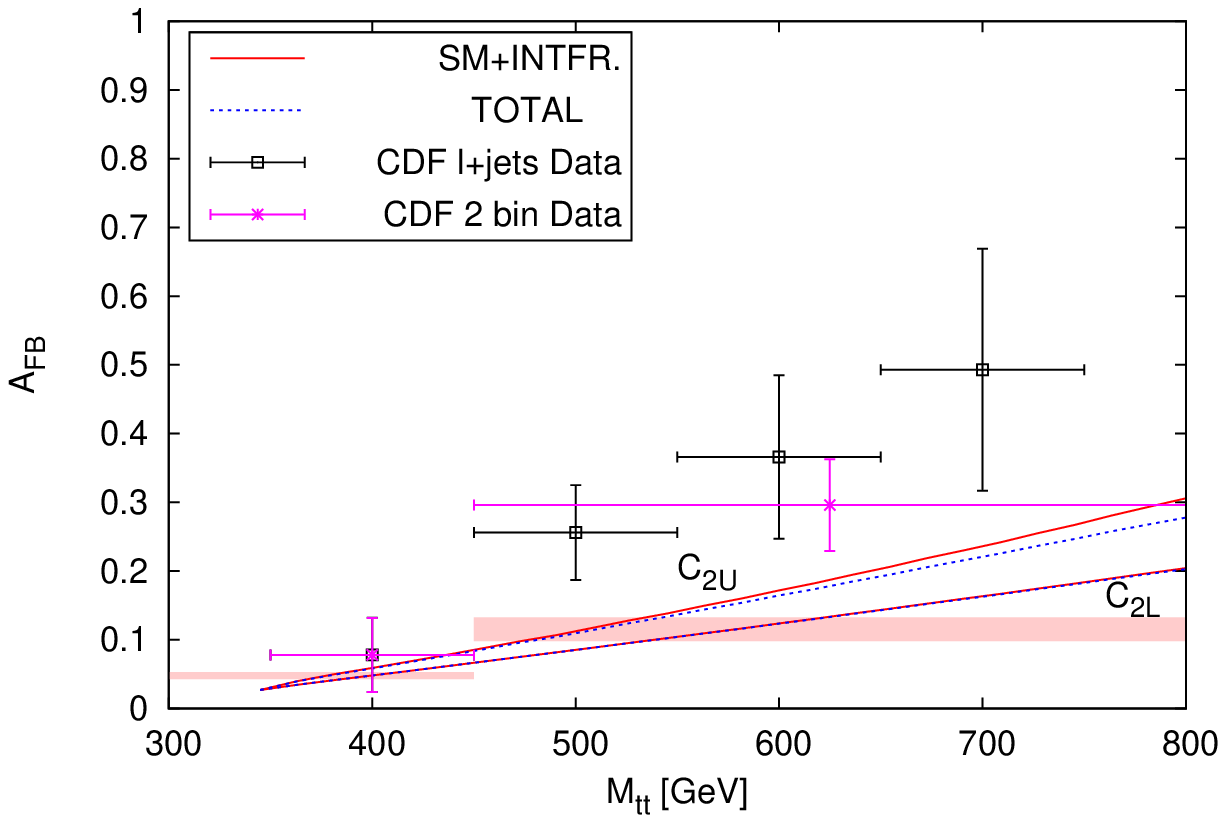,height=4.0cm,width=7.0cm}}\\
{\epsfig{figure=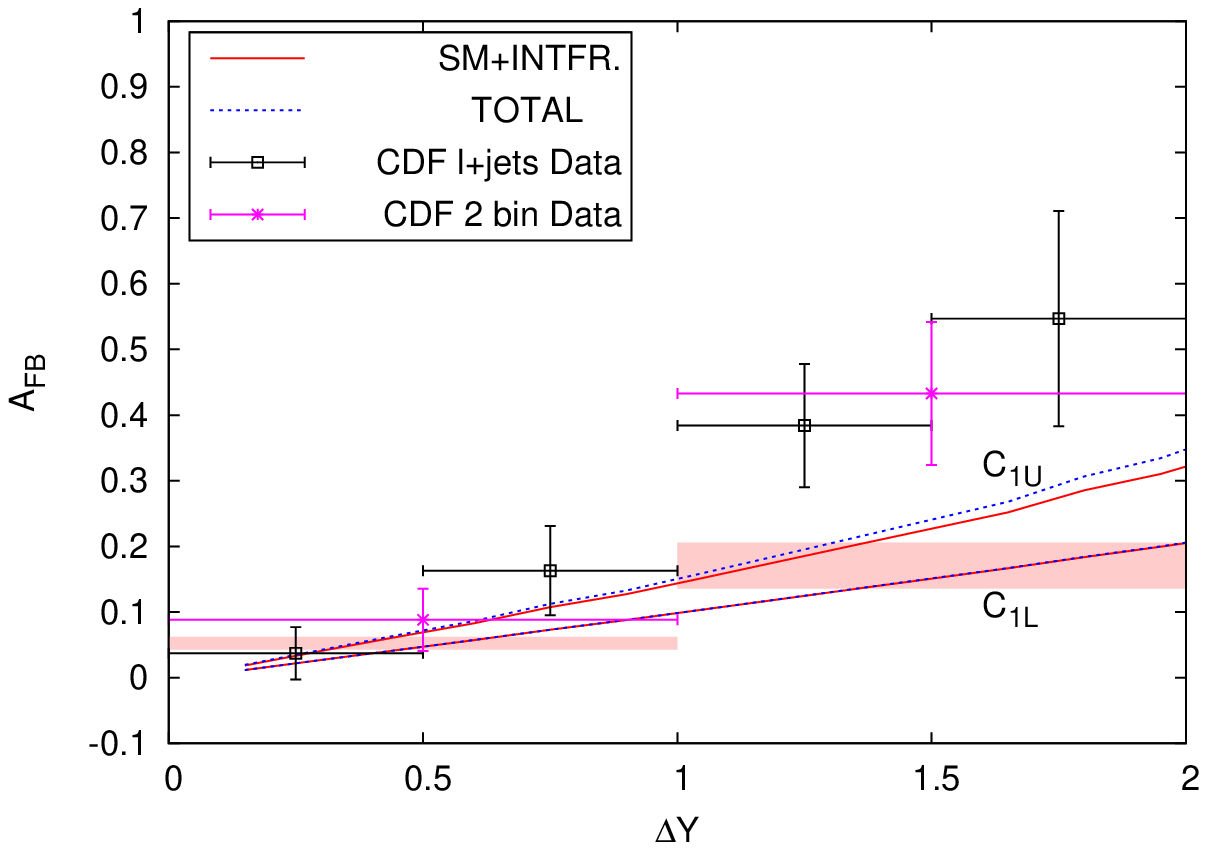,height=4.0cm,width=7.0cm}} &
{\epsfig{figure=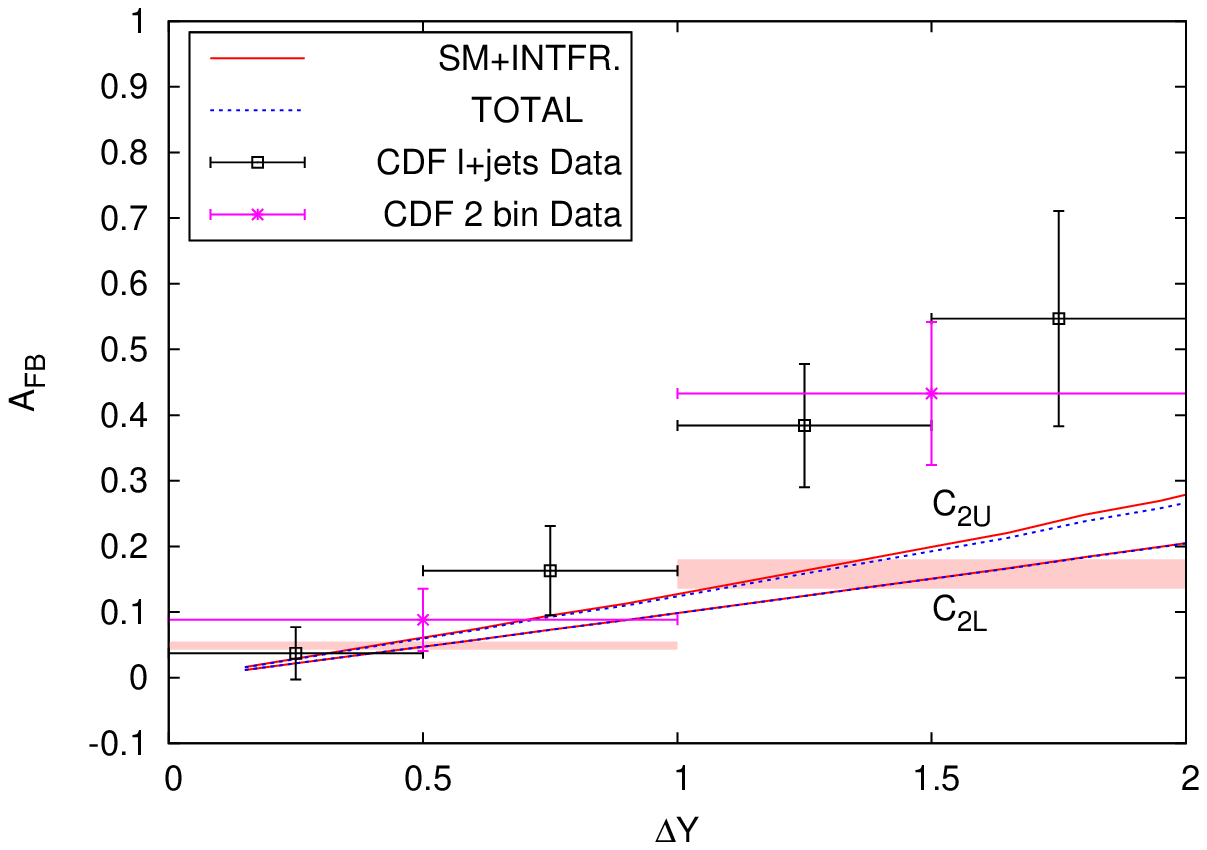,height=4.0cm,width=7.0cm}}\\
\end{tabular}
\end{center}
\vspace{0.0cm}
\caption{\it 
Top FB asymmetry as functions of $M_{t\bar{t}}$  (upper)
and $\Delta y$ (low). In the left frames we are taking
$C_1$ in the range between
$C_{1L}=0.15$ and $C_{1U}=0.97$ with $C_2=0$.
In the right frames, we vary $C_2$  in the range between
$C_{2L}=-0.15$ and $C_{2U}=-0.67$ with $C_1=0$. 
We have taken $\Lambda = 1$ TeV in both cases.
In each frame, the two bands are for  $A_{\rm FB}$ in the
lower and higher $M_{t\bar{t}}$ or $\Delta y$ bins
varying $C_1$ (left) and $C_2$ (right)
in the ranges delimited by $C_{1L,1U}$ and $C_{2L,2U}$, respectively,
and the dots for the CDF data with errors.
In the solid (red) lines, we 
include only the SM contribution and
the one from the interference between the SM and NP amplitudes while
the effects of $(NP)^2$ term have been added
in the dotted (blue) lines.
}
\label{fig:afb}
\end{figure}

Now we are in the place to present the numerical results. In the first place, we consider
the case either of $C_i$ is zero. It is reasonable point in the parameter space 
since many of the models predict
only one of Wilson coefficients are nonzero. See the Table(1) of Ref. \cite{Jung:2009pi}. 
The allowed ranges by the integrated FB asymmetry then become $C_1 \sim [~0.15, 0.9~]$ and
$C_2 \sim [~-0.67, -0.15~]$ with $C_2$ or $C_1$ is zero for either case. The results are 
presented in Fig.~\ref{fig:afb}. One can see that for models that produce nonzero values 
 for either $C_1$ or $C_2$ it is impossible to fit all data altogether. They fails to 
generate right behaviors of FB asymmetry as functions of invariant mass and rapidity. They 
cannot fit the two-bin data either though the points satisfy the integrated asymmetry. So 
we can conclude that such models can be excluded if we trust the Tevatron observation in 
the sense that they fail to reconstruct the consistent prediction compared to data.

 Next, let us consider the more general case of nonzero values for both $C_1$ and $C_2$. 
We assign the constraint that $C_1+C_2=0$, which makes no contribution to the cross section.
(See Eq. (\ref{eq:bsm}) ) It is reasonable assumption 
since the variation of the cross section is smaller than that of $A_{\rm FB}$. 
In addition to that, we further assume that $C^{RR}_{8q}=C^{LL}_{8q}=\frac{1}{2} C_1$ and 
$C^{RL}_{8q}=C^{LR}_{8q}=-\frac{1}{2} C_1$ to minimize the the $NP^2$ contribution which 
spoil the validity of the effective Lagrangian approach. 

\begin{figure}[t]
\begin{center}
\hspace{-0.0cm}
\epsfig{figure=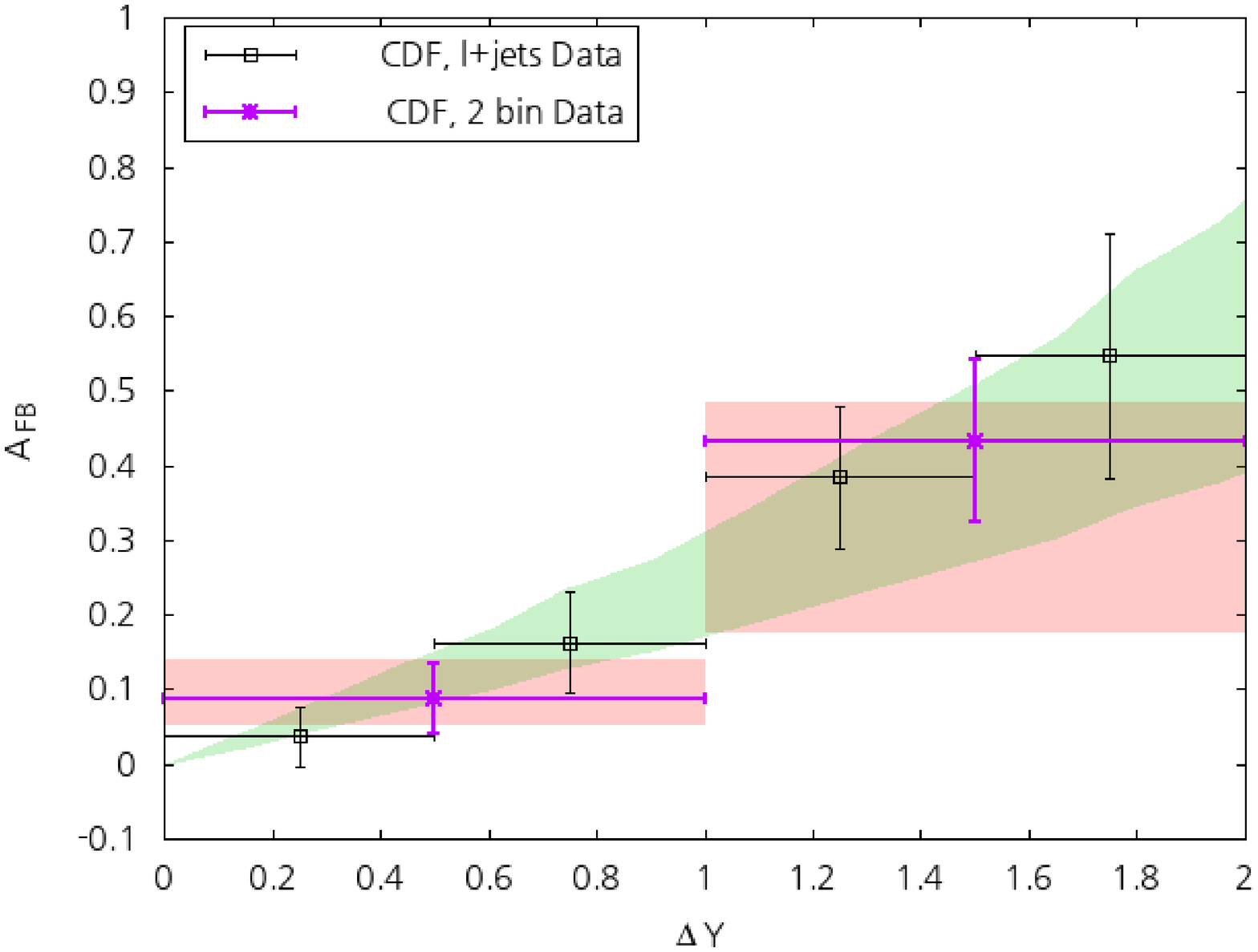,width=7cm,height=4.0cm}
\epsfig{figure=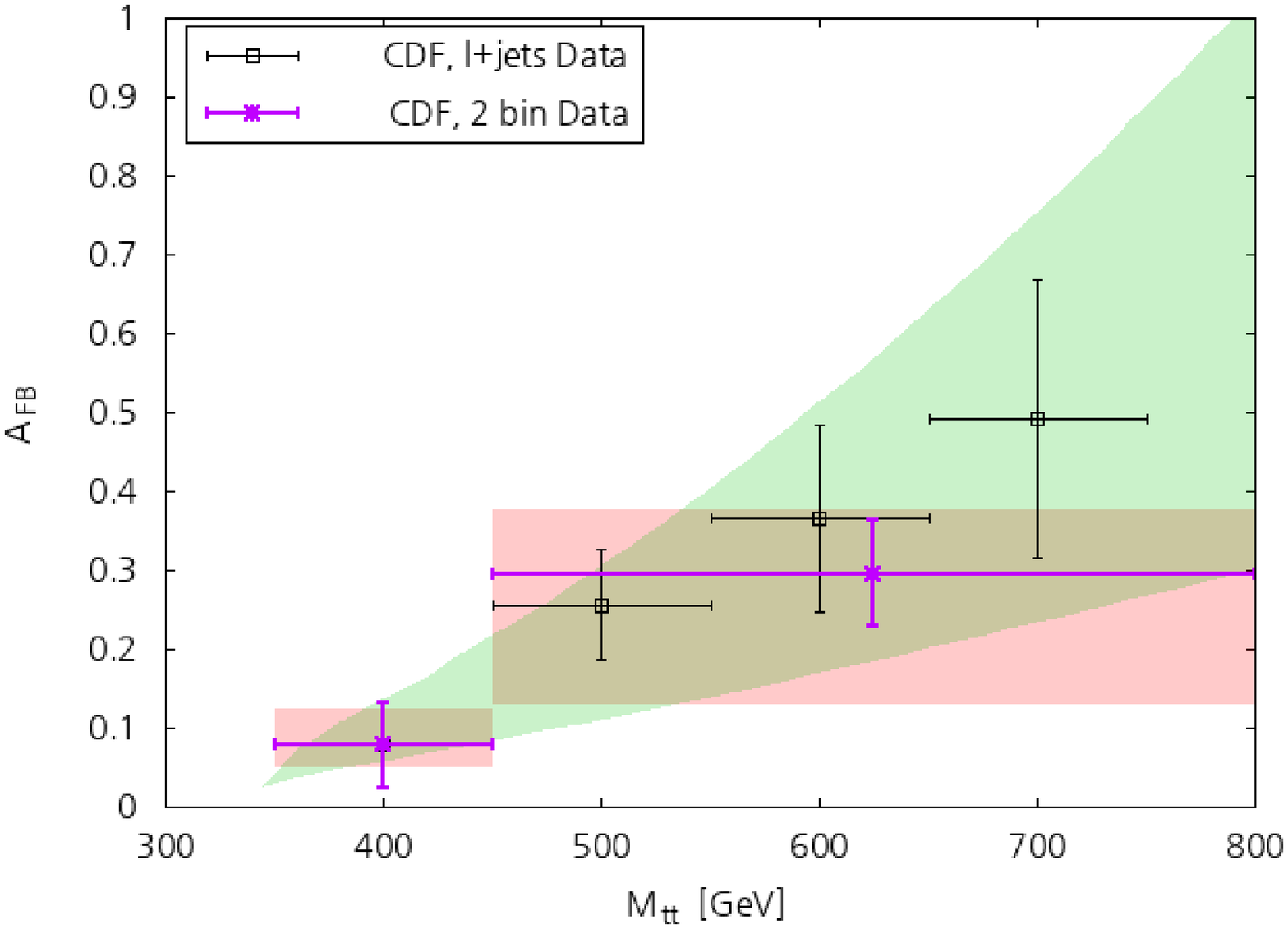,width=7cm,height=4.0cm}
\end{center}
\caption{\it Same as Fig.(\ref{fig:afb}), with $C_1+C_2=0$. 
Experimental results are added in the figures in the same ways.}
\label{fig:afb2}
\end{figure}

The results are collected in Fig.(\ref{fig:afb2}). Unlike the previous results, now we can fit the data. 
Actually, this parameter set corresponds to usual axigluon models. As shown in the figure,
experimental results are fit in consistent way this time. Interesting point is that this holds for 
rather larger values of integrated asymmetry. In other words, if we restrict the integrated 
asymmetry in the lower half range of 1-$\sigma$ result, then the model cannot be accommodated 
to the observations. The parameter set which provide the consistent values for 
all the variables prefers the larger value of integrated asymmetry, roughly $A_{\rm FB} 
\sim [0.158, 0.158+0.074]$, to the smaller values of integrated asymmetry, namely, 
$A_{\rm FB} \sim [0.158-0.078, 0.158]$. 
The validity of this assertion can be tested if more experimental results are accumulated. 
Unfortunately it cannot be done and we only can tell about rather vague tendency instead of 
giving strict statements. 
Conservatively speaking, we can assert that the parameter set of $C_1+C_2=0$ fit the data 
well with preference for the larger value of integrated asymmetry. 

Final comments are in order. We should always be careful for treating the problem from the
effective Lagrangian method. In that case we should make it sure that the higher order 
contributions are small enough not to spoil the validity of the 
nonrenormalizable operators. For that purpose we estimate the relative size of the
$NP^2$ over the $SM$ value. The ratio goes from $0.08 \%$ up to at most 35$\%$ 
of the $SM$ value in the large invariant mass region, and we can state that 
the analysis remains robust in most range. Similar behavior takes place in the 
rapidity dependent asymmetry.


\section{Summary}
We perform the analyses on the invariant mass and rapidity dependent top forward-backward 
asymmetries from the effective Lagrangian viewpoint. With the non-observation of any 
new resonances at the Tevatron, such method is efficient way to extract the new physics 
information that might be responsible for FB asymmetry. We show that generically the models
 that give the nonzero value for only one of $C_1$ and $C_2$ cannot reproduce the 
experimental observations in consistent way in the sense that they fail to fit the slope 
and two-bin values. If we relax the constraint and assign the axigluon-like condition as
$C_1+C_2=0$ then we can fit the data. Even in that case large integrated FB asymmetry are
disfavored. More study including the LHC data may be necessary to clarify the experimental 
status of various model in relation with the information extracted by the effective 
Lagrangian approach at the Tevatron.

\section*{Acknowledgements}
This work is supported in part by Basic Science Research
Program through NRF 2011-0022996 and in part by NRF
Research Grant 2012R1A2A1A01006053.

\end{document}